\documentstyle[aps,preprint,epsfig]{revtex}

\preprint{ECT* 05-2000}
\title{\bf Finite formation time effects in quasi-elastic $(e,e')$
 scattering on nuclear targets}
\author{Mikhail A.\,Braun}

\address{ Department of High-Energy Physics,
S.Petersburg University, 198904 S.Petersburg, Russia}
\author{Claudio\,Ciofi degli Atti}

\address{ Department of Physics, University of Perugia, and Istituto Nazionale
di Fisica Nucleare, Sezione di Perugia, Via A. Pascoli, I--06100 Perugia, Italy}
\author{Daniele\, Treleani}

\address{ Department of Theoretical Physics, University of Trieste, Strada
Costiera 11, Istituto Nazionale di Fisica Nucleare, Sezione di Trieste,
and ICTP, I--34014, Trieste,Italy}

\def\beq{\begin{equation}}
\def\eeq{\end{equation}}
\def\noi{\noindent}
\begin{document}
\maketitle
\medskip
\vspace{1 cm}
{\bf Abstract}

The problem of the  final state interaction, in
quasi-elastic $(e,e')$ scattering at large $Q^2$, is investigated
by exploiting the idea that the ejected
nucleon needs a finite amount of time to assume its asymptotic form.
It is shown that  when the dependence of the
scattering  amplitude of the ejected nucleon on its virtuality
is taken into account, the
final state interaction is  decreased.
The developed  approach is simpler to
implement than the  one based on the color transparency
description of the damping of the final state interaction, and is
 essentially equivalent
to the latter in the case of the  single rescattering
term. The $(e,e')$ process on the deuteron is numerically
investigated and it is shown that, at $x=1$,  appreciable
finite formation time effects at $Q^2$ of the order of 10 (GeV/c)$^2$
are expected.

\section{Introduction}
Quasi-elastic (q.e.) $(e,e')$ scattering on nuclei  is
considered to be  a suitable process
 to look for color transparency (CT) effects in QCD [1,2].
The original idea was that at high $Q^2={\bf q}^2-{\nu}^2$ ( ${\bf q}$ and
${\nu}$ being the three-momentum and energy transfers, respectively)
the state which
emerges after the interaction ("the ejectile state") is dominated by
 configurations
of  small
size  $\rho\sim \sqrt{1/Q^2}$.
Since the color charge is supposed to be neutralized at
small distances , the final state interaction (FSI) of the
ejectile should vanish at high $Q^2$, providing, by this way,  a clear
signature of the underlying production mechanism.
Unfortunately,  a more detailed analysis reveals that the situation is not
that simple,  for it can be trivially shown
that the transverse dimension of the
ejectile is exactly equal to the one of the initial struck nucleon;
this fact
however is not, in principle, in  contradiction with the vanishing of FSI
at large $Q^2$ [3,4].

The latter  effect, if operative, is a consequence
of the cancellation between the various
contributions of the different
intermediate states produced after the initial interaction, in
particular, of those with a large mass. In order to have a detailed theoretical
description of CT,  one should then be able to describe
the propagation through the nucleus of all possible states of the ejectile,
including the ones with very high masses, with
the vanishing of FSI  resulting from
the destructive interference of all these  different contributions.
The practical implementation of such a program looks therefore  as a
rather difficult task.
Summing over all excited ejectile states seems to be technically feasible
only in $3q$ or quark-diquark oscillator models, for which CT
does not occur. The authors using these
models are in fact forced to impose artificially CT by introducing
a transverse form-factor $\sim \exp (-\rho^2Q^2)$ [5,6].
Moreover, the high-mass states of the ejectile cannot
certainly be described in terms of such models and require the inclusion
of gluons and sea quarks.

In this note we discuss a different approach to the problem.
We want in fact to take into account the
finite formation time (FFT) of the finally observed proton,
which is an alternative and possibly
more convenient way to represent the vanishing of FSI in
q.e. $(e,e')$ process at high $Q^2$.
Let us first of all recall some relevant features of the theoretical
description of high energy hadron-nucleus interaction.
It has long been
known that after a particle has undergone an interaction, it
 should elapse a certain amount of time before it  becomes
  capable of a new one. Such a phenomenon, is formally due to
    the vanishing of the
contribution of all planar diagrams at high energies. By studying the
planar diagrams, one can ascribe  the mechanism responsible for such an effect
to the  cancellation of the  contributions to the absorptive part  coming
from different intermediate states. From  a more formal point of view,
it results from the dependence on the virtualities of the off-shell
amplitudes appearing in the diagrams with several interactions.
In the FFT approach,  multiple interactions correspond to the following
picture:  the ejectile  splits into its components
(partons) which then interact with the target in parallel. From the
diagrammatic point of view, this contribution is represented by non-
planar diagrams which substitute the planar ones as energy increases.
Note that in the dispersion approach by  Gribov [7] this substitution
is not felt at all. One just closes the contour of integration over
cumulative momentum transfers around the right-hand singularities.
By doing so, one however assumes that there are also some non-zero
left-hand singularities: otherwise the result would have been zero.
This is precisely what happens if one takes only planar diagrams into account:
they have no left-hand singularities. Then, although
closing the contour around the right-hand singularities one
seemingly obtains some non-zero contributions (specifically from
the pole corresponding to the propagating particle itself), the sum
of all contributions is finally equal to zero.

In  q.e. $(e,e')$ scattering on nuclei the situation is rather
different, for
the space-time point of the creation of the hadronic state, after the
interaction of the photon  with a nucleon as a whole,
is fixed and it is inside the
nucleus. Thus, unlike the case of hadron-nucleus interaction, where the
projectile interacts while being in its asymptotic state, the hit
nucleon  will become capable of  a new interaction
only after the
formation time which grows with its velocity. At large $Q^2$ it
will then be able to interact only outside the nucleus
and all FSI will vanish.
In the diagrammatic language, the rescattering of the ejectile always
includes a planar diagram. In the dispersion approach this means that
there are no left singularities in the cumulative momentum transfers.
Therefore one can  expect that the FSI in   q.e.
$(e,e')$ scattering  will  die out at high $Q^2$.

It should be pointed out  that the mechanism which makes the FSI vanish
due to CT or FFT is the same:
the cancellation of the  contributions from various propagating
states. Many authors believe  that in QCD the FFT
is a direct consequence of  color transparency: a colorless quark system
created at a point needs a finite time to reach its asymptotic configuration
with the corresponding cross-section [8-10]. The FFT looks however as a
more general property, whose origin is not in the color dynamics of QCD, but
is rather related to the vanishing of the planar diagrams, that can
be demonstrated to occur  at high energy for the so-called soft field theories,
of the $\lambda\phi^3$ type, and that is expected to hold also in QCD at
small $x$[11].

The FFT  concept allows one to develop  a description of FSI effects in
 q.e. $(e,e')$ scattering,
which explicitly fulfills the requirement of their vanishing  at high $Q^2$,
so that all the necessary cancellations
are automatically implemented.
In the present paper we discuss the  possible modifications of
the standard Glauber formula,
allowing the amplitudes to depend on their
virtualities in a way compatible with the standard analytical properties
so as to guarantee the vanishing of the amplitudes at high virtualities,
which is precisely the property which leads to all
FFT effects. It should be noticed that our approach is not very much
  more restrictive
than the currently used multi-channel Glauber approach. In fact,
a virtuality dependent nucleon-nucleon amplitude
simulates, to a large extent, the
propagation and the interaction of excited intermediate
states.
Thus a model for the
propagation of a $3q$ state through the nucleus can be
(approximately) translated into
the propagation of the nucleon with a particular dependence
of its interaction amplitude on the virtuality. The case of the
single  rescattering
 will be worked out in detail, showing
that in this case
the two approaches are equivalent. In the FFT approach,
model building means specifying the dependence of the amplitude on the
virtuality,
which phenomenologically seems to be a simpler
task than constructing a model for the propagation and the interaction
of  excited
$3q$ states, as required by the  conventional description of color transparency.
The present approach can be properly generalized so as to take into
account also the excited nucleon states.
We do not try to discuss  such a  more elaborate feature in the present paper,
and we assume,  moreover, the simplest possible  factorizable  form of
dependence of the amplitudes on their virtualities. Our  final result
is that
final state interactions vanish at high $Q^2$ as $mM^2R_A/Q^2$,
where $R_A$ is the radius of the nucleus, $m$ the nucleon mass and
$M^2$ its  average excitation mass squared.
The case of FSI with a deuteron target will be considered in detail (an interesting
approach to the problem has also been discussed in ref.[12]).
Although
in this case one  does not expect appreciable FSI effects,
 the two-body system has the advantage that its structure is well
  known, and, moreover, only the single rescattering term has to be
  considered, in which case,
   as it will be shown,
 different theoretical approaches to FSI converge to the
same result.

Our  paper is organized as follows. In Section 2 the  general
formalism for treating the FSI in the Glauber approach when the  amplitudes
 depend
on the virtuality of the external lines is presented;
in Section 3
 the high $Q^2$ limit of the approach is investigated; Section 4 is devoted
  to the numerical application to the deuteron
target; finally, the conclusions are drawn in
Section 5.

\section{Formalism}

A straightforward way to incorporate  FFT effects, generated by the
dependence of the scattering amplitudes on the virtualities of the
colliding particles, is through the
Feynman diagrams formalism.
 The amplitude describing $n$ consecutive rescattering of
the ejectile
emerging from the interaction of the struck nucleon  with the
incoming virtual photon, is depicted in Fig.1. It corresponds to the usual
 Glauber approximation
of the scattering amplitude [13].
Our notations are as follows:
{\it i})
The four-momentum of the target nucleus is
denoted  $Ap$ and
we work in rest system of the nucleus, so that ${\bf p}$=0;
{\it ii}) the momenta of the nucleons before (after) all interactions
 are denoted $k_i$
($k'_i$); {\it iii}) the spectators correspond to $i=n+2,...A$, for which
$k_i=k'_i$;
{\it iv}) the active nucleon is labeled $1$, and  nucleons from 2 to $n+1$
are the ones
on which the active nucleon $1$ rescatters. Correspondingly,
the number of  rescatterings goes  from 2 to $n+1$; {\it v}) the
momentum transferred in the $i$th rescattering is $q_i$, so that
$k'_i=k_i+q_i$ for $i=2,...n+1$; {\it vi})the momentum of the active nucleon
after interaction with the photon is $k_1^{(1)}=k_1+q$, and after
 the $i$th
rescattering, $k^{(i)}_1=k_1+q-\sum_{j=1}^i q_j$.

The  derivation of the amplitude corresponding to Fig. 1 is a  standard
one, so that, in the following,
 only those  points which are related to the FFT
effects, generated by  the
dependence
of the amplitudes on the  virtuality of the nucleons, are stressed; more details
on  the derivation of the amplitude  are given in the Appendix.

The expression for the amplitude with $n$ rescatterings, corresponding
to Fig.1 reads as follows
\[
i{\cal A}^{(n)}=(1/2)xQ^2(4m\nu)^n\frac{A!}{(A-n-1)!}
\int\prod_{j=2}^{n+1}\frac{d^4k_j}{(2\pi)^4}
\frac{d^4k'_j}{(2\pi)^4}P(k_j)P(k'_j)\prod_{j=n+2}^{A}\frac{d^4k_j}
{(2\pi)^4}P(k_j)\]\beq
\times P(k_1)P(k'_1)P(k_1^{(1)})\prod_{j=2}^{n+1}P(k^{(j)}_1)if_j
i\gamma(k_1,q) i\gamma(k'_1,q)i\Phi(k_i)i\Phi(k'_i)
\eeq
In Eq.(1),
$P(k)$ denotes  the propagator of the nucleon with momentum $k$, and
$f_j$, $j=2,..n+1$ the corresponding  rescattering amplitudes, which
 are assumed to
depend,  besides
 the energy and momentum transfer, also upon  the virtuality of the fast nucleon
 which has been struck by the photon.
On the mass shell, they are normalized according to
\beq
2{\Im}{\rm m}f=\sigma^{(tot)}
\eeq
(the factor $(4m\nu)^n$ originates from this normalization).
 The vertex $\Phi$ describes the transition
of the nucleus into $A$ nucleons and
the vertex $\gamma$ is the form-factor of the on shell nucleon
( only
the electric form-factor  appears in  our spinless model).
The factor $(1/2)xQ^2$ originates from the initial photon interaction, and
$x={Q^2}/(2k_1\cdot q)$ is the usual Bjorken scaling variable.
With our normalization,
twice the imaginary part of the amplitude gives the
corresponding contribution to the structure function $F_2$.

The integrations over the  energies and the transverse components of all momenta,
as well as over the longitudinal momentum components of the spectators, are
performed  in the
standard way. As a result,  the amplitude is expressed
as an integral over the transverse coordinate $b$ of the struck nucleon
with respect to the direction of the virtual photon
and  the longitudinal
interaction
points $z_i$, $i=1,2,...n+2$, with the appropriate nuclear
density matrices.
One is left with the non-trivial momentum integrations over the $n+1$ $z$-
components of $k_1$  and $q_i$, $i=2,3,...n+1$, through which the
virtualities of
the active nucleon are expressed as
\beq
v_i=(k_1^{(i)})^2-m^2=Q^2\left(\frac{1}{x}-1\right)+\frac{Q^2}{xm}
\left(\sum_{j=2}^i q_{jz}-k_{1z}\right)
\eeq

For the sake of simplicity we make the simplest possible assumption
on the dependence of the amplitude
$f_j$ on the two virtualities $v_j$ and $v_{j-1}$, namely we assume the
factorized form
\beq
f_j=F(v_{j-1})F(v_j)f
\eeq
where $f$ is the on-shell amplitude and $F(v)$  a form-factor
 exhibiting the dependence of $f$ on the virtuality of
the external lines,
normalized according to $F(0)=1$ and decreasing with $v$.  In the same manner we
introduce also the dependence of the off-mass-shell electric
form-factor on the virtuality of the nucleon
\beq
\gamma(k_1, q)=F(v_1)\gamma(Q^2);\ \ \gamma(k'_1,q)=F(v_{n+1})\gamma(Q^2)
\eeq
The whole $v$-dependence of the integrand
is then given by the factorized expression
\[\prod_{i=1}^{n+1}\frac{F^2(v_i)}{-v_i-i0}\]
multiplied by the exponential
\[
\exp(i\Delta (z_{n+2}-z_1)),\]
which does not depend on $v_i$,
and  the exponential
\[
\exp\left( i\frac{xm}{Q^2}\sum_{j=2}^{n+1}v_j(z_j-z_{j+1})\right)
\]
which is symmetric in all $v_j$'s. In these expressions we have
used $\Delta=m(1-x)$ and have denoted
$z_{n+2}\equiv z'_1$ (the last interaction point in the longitudinal
space).
All  integrations on the virtualities $v$ lead therefore to the same function
\beq
iJ(-z)=\int\frac{dv}{2\pi}\frac{F^2(v)}{-v-i0}\exp\left(i\frac{xm}{Q^2}vz\right)
\eeq

If nuclear correlations are disregarded, the vertex functions
$\Phi$ can be expressed through the factorized  nuclear density
matrix

\beq
\rho(bz_1,bz_2,...bz_{n+1}|bz_{n+2},bz_2,...bz_{n+1})=
\rho(bz_1|bz_{n+2})\prod_{j=2}^{n+1}\rho(bz_j)
\eeq
where (cf. Fig.1) the above quantity is non diagonal in the coordinate
$1$, diagonal in coordinates $2$...$(n+1)$, and it is integrated
over the coordinates $(n+2)...A$.
As a result we obtain  the amplitude
with $n$ rescatterings in the following form
\[
{\cal A}^{(n)}=\gamma^2(Q^2){f}^n
\frac{x^2m}{2}\frac{A!}{(A-n-1)!}\]\beq
\int d^2bdz_1dz_{n+2}\prod_{j=2}^{n+1}dz_j\rho(bz_j)
\prod_{j=1}^{n+1}iJ(z_{j+1}-z_j)
\exp (i\Delta(z_{n+2}-z_1))\rho(bz_1|bz_{n+1})
\eeq

With the dependence of the amplitudes on the virtualities turned on,
all effective propagators $J(z)$ go to zero in the large $Q^2$
limit. Thus,  for large $Q^2$, the bulk of the rescattering contribution
comes from the single rescattering term
\beq
{\cal A}^{(1)}=\gamma^2(Q^2)f
\frac{x^2m}{2}A(A-1)
\langle\int dz\rho(b_1,z)
iJ(z'_1-z)iJ(z-z_1)\rangle_1
\eeq
where the notation
 $\langle...\rangle_1$ means that the quantity in brackets, i.e.
 $O(b_1,z_1',z_1) \equiv \int dz\rho(b_1,z)
iJ(z'_1-z)iJ(z-z_1)$, has to be averaged over  the
coordinates of the active nucleon, according to
\beq
\langle O(b_1,z_1',z_1)\rangle_1\equiv\int d^2b_1dz_1dz'_1\rho(b_1z_1|b_1z'_1)
\exp(i\Delta(z'_1-z_1))O(b_1,z_1',z_1)
\eeq

If the dependence of the $f$'s on their virtualities is neglected and
one
puts $F(v)=1$,
then $J(z)=\theta(z)$ so that, after summing over $n$ in Eq.(8),
the standard Glauber result is obtained
\beq
{\cal A}=iA(1/2)mx^2\gamma^2
\langle
\left((1+if
T(b_1,z'_1,z_1))^{A-1}-1\right)\rangle_1
\eeq
where
\beq
T(b_1,z'_1,z_1)=\int_{z_1}^{z'_1} dz \rho(b_1,z)
\eeq

The introduction of  the dependence of the amplitudes
on the virtualities,  by means of the
factorized approximation (4), is effectively equivalent to the
replacement of  the usual
$\theta(z)$
in the nucleon
propagator with the function
$J(z)$, which depends on the virtuality through the  form-factor $F(v)$.

This simple rule suggests a possible different derivation of the scattering
amplitude, which is equivalent to
the one just considered,  when the dependence of the amplitude on
the virtuality of the external lines is turned off.
 Instead of choosing a particular set of Feynman diagrams, one may assume that
the scattering matrix on the nucleus factorizes
into the product of scattering matrices on individual nucleons.
In our case we assume that the scattering matrix of the 1st (active)
nucleon on the other  $A-1$ nucleons is given by
\beq
S(r_2,r_3,...r_{A}|r_1)=\prod_{j=2}^{A}s(r_j)
\eeq
where $r_j=(b_j,z_j)$, and $s(r_j)$ is the  scattering matrix on the nucleon $j$.
In accordance with the results obtained from the consecutive
rescattering diagram we take
\beq
s(r_j)=1-J(z'_1-z_j)J(z_j-z_1)\Gamma(b_j-b_1)
\eeq
where  $b_1$ is the impact parameter of the active nucleon, $z_1$ and $z'_1$
its longitudinal coordinates before and after the interaction, and $\Gamma$
the nucleon-nucleon profile as a function of the relative transverse distance of
the two interacting nucleons, normalized as $\int d^2b\Gamma(b)=-if$. The
two functions $J$, which  describe the  propagation of the active nucleon
to and from the collision
point $z_j$,
 are replaced by $\theta$ functions in the standard Glauber approach.
By averaging (13) over the positions of the nucleons from 2 to $A$,
by making
use of the
approximation given by  Eq.(7) and by disregarding the dimensions of the
nucleon as compared with the dimensions of the nucleus, we obtain
\[
\langle S(r_2,r_3,...r_{A}|r_1)\rangle_{A-1}\equiv
\int\prod_{j=2}^A d^3r_j\rho(b_jz_j)S(r_2,r_3,...r_{A}|r_1)
\]\beq=
\left(1+if\int dzJ(z'_1-z)J(z-z_1)\rho(b_1,z)\right)^{A-1}
\eeq
The final amplitude is obtained  by averaging Eq.(15) over the coordinates
of the 1st nucleon and by multiplying the result with the appropriate factor,
as in Eq.(11).

It can be readily seen that,
 in absence of any dependence upon the  virtualities, that is, when
$J(z)$ is replaced by  $\theta(z)$, the standard  Glauber result (11)
is recovered.
However it is instructive to notice that, with the dependence
on virtualities turned on, the expression (15) is generally different
from (8) obtained from the consecutive scattering diagram.
Only the single rescattering contribution is identical in (8) and (15).
Thus the assumption of the factorizability of the
nuclear scattering matrix (13) generally has a physical
meaning different from selecting the consecutive scattering diagrams.
The factorizability (13) seems to be a more fundamental assumption,
applicable also to high energy scattering. As we shall see,
the production amplitudes  derived on its basis reproduce correctly the
Glauber result in absence of any dependence on the virtualities, and
thus satisfy the AGK rules[14], in contrast to the amplitudes obtained from
the discontinuities of (8) (see Appendix). The consecutive scattering diagram
approach to FSI, with the factorization hypothesis (4), gives moreover
problems with unitarity, which are avoided when FSI is expressed as in
Eq.(15). In the following we will therefore work in the scheme
where the $S-$matrix is factorized as a
product of scattering matrices on individual nucleons. The two approaches give
in any case the same result for the deuteron,
which will be considered in sec. IV.

We are interested in q.e. $(e,e')$ scattering on nuclear targets, i.e. in
the process involving the
 production of a proton in the final state.
In absence of any FSI, the contribution of this process to the inclusive
structure
function
$F_{2}^{N/A}$, to be denoted $F_{2}^{N/A,(0)}$, is proportional to the
 square modulus of the nucleon
photoproduction amplitude and it is
given by
\beq
F_{2}^{N/A,(0)}(x,Q^2)=(1/2)A\gamma^2(Q^2)x^2m\langle 1\rangle_1
\eeq
The FSI is obtained by  multiplying each production amplitude
by the $S$ matrix (13) taken between
the initial and final nuclear states. Since the final active nucleon
is physical, its virtuality is zero and, accordingly, one has  $J(z_1'-z_j)=1$.
The scattering
matrix on a nucleon
should then contain only one function  $J$:
\beq
s(r_j)=1-J(z_j-z_1)\Gamma(b_j-b_1)
\eeq

After summing over the final nuclear
states one obtains
\beq
\langle S(r_2,r_3,...r_{A}|r_1) S^*(r_2,r_3,...r_{A}|r'_1)\rangle_{A-1}
\eeq
which replaces $\langle 1\rangle_1$ in the average of Eq.(16).
Eq.(18), which can be evaluated  in a straightforward way, differs from
the
Glauber result  [16] only by the substitution of all functions
$\theta (z)$ with
the functions $J(z)$. One thus obtains
\[
\langle S(r_2,r_3,...r_{A}|r_1) S^*(r_2,r_3,...r_{A}|r'_1)\rangle_{A-1}
\]\[=
\Big(1+if\int dz J(z-z_1)\rho(b_1,z)-if^*\int dz J^*(z-z'_1)\rho(b_1,z)
\]\beq+
\sigma^{el}\int dzJ(z-z_1)J^*(z-z'_1)\rho(b_1,z)\Big)^{A-1}
\eeq

From Eq. (19) one obtains the single rescattering contribution to
 the inclusive structure function in the following form
\[
F_{2}^{N/A,(1)}=
(1/2)A(A-1)\gamma^2(Q^2)x^2m\]\beq
\langle\int dz
\rho(b_1,z)
(ifJ(z-z_1)-if^*J^*(z-z'_1)
+\sigma^{el}J(z-z_1)J^*(z-z'_1))\rangle_1
\eeq
The expression above can  also be obtained from the discontinuities of the
amplitude (8), corresponding to the consecutive rescattering diagram of Fig.1
(see Appendix).

\section{Unitarity and high $Q^2$ behavior}
At high $Q^2$ the dependence of the
amplitudes on their virtualities becomes of primary importance.
On rather general grounds we may express the form-factor
squared as
\beq
F^2(v)=\int_0^{+\infty}\frac {dv'v'\tau(v')}{v'-v-i0}
\eeq
with the normalization
\[
\int_0^{+\infty} dv\tau(v)=1
\]
Note that $\tau(v)$ needs not be real.
From (6) we find
\beq
J(z)=\theta(z)\int_0^{+\infty}
dv\tau(v)\left(1-\exp\left(-i\frac{xmvz}{Q^2}\right)\right)
\eeq

The simplest choice of $\tau(v)$ is evidently
$\tau(v)=\delta(v-M^2)$ and in this case
\beq
J(z)=\theta(z)
\Big[1-\exp\left(-i\frac{z}{l(Q^2)}\right)\Big]
\eeq
where
\beq
l(Q^2)=\frac{Q^2}{xmM^2}
\eeq
has the obvious meaning of a formation length
growing linearly with $Q^2$.

In this case, at the single rescattering level,
our model for the FFT
 coincides with the standard two-channel Glauber model
for the propagating nucleon and its excited state of mass squared
${m^*}^2=m^2+M^2$, provided that the amplitudes and the production vertices
are constrained in a definite way.
Indeed, using (23)in  Eq.(9), we obtain
\[
{\cal A}^{(1)}=-\gamma^2(Q^2)f
\frac{x^2m}{2}A(A-1)
\langle\int dz\rho(b_1,z)\theta(z'_1-z)\theta(z-z_1)
\Big(1-e^{-i(z'_1-z)/l}\]\beq-
e^{-i(z-z_1)/l}+e^{-i(z'_1-z_1)/l}\Big)
\rangle_1
\eeq
On the other hand, the two-channel Glauber model with two
ejectile states 1 (the nucleon) and
2 (its excited state) leads to the single rescattering contribution
\[
{\cal A}^{(1)}=-\frac{x^2m}{2}A(A-1)
\langle\int dz\rho(b_1,z)\theta(z'_1-z)\theta(z-z_1)
\Big(\gamma_1^2f_{11}+\gamma_1\gamma_2f_{21}e^{-i(z'_1-z)/l}
\]\beq+
\gamma_1\gamma_2f_{12}e^{-i(z-z_1)/l}+
\gamma_2^2f_{22}e^{-i(z'_1-z_1)/l}\Big)
\rangle_1
\eeq
where $f_{ik}=f_{ki}$, $i,k=1,2$ are the forward scattering amplitudes for
transitions $i\rightarrow k$ and $\gamma _i$, $i=1,2$ are vertices for the
production of the two ejectile states.  One immediately observes that
(25) and (26) coincide if
\beq
f_{11}\gamma_1+f_{12}\gamma_2=0,\ \
f_{21}\gamma_1+f_{22}\gamma_2=0
\eeq
and, moreover,  if  $\gamma_{11}^2f_{11}$ in (26) is identified
 with $\gamma^2f$ in (25).
The meaning of the sum rules (27)
is that when applying the matrix $f_{ik}$ to the vector $\gamma_i$ one obtains
zero, which is the condition for propagating eigenstates of the forward
scattering matrix with zero eigenvalue in nuclear medium [16]. As
discussed
in ref.[17], it is precisely the condition for color transparency.

In fact in the case of two channels one may easily see that both unitarity,
$2{\Im}{\rm m}f_{il}=\sum_{j=1,2}f_{ij}f^*_{jl}$, and the transparency
conditions are
satisfied
by
\beq
f_{12}=f_{21}=-\xi f_{11},\ \
f_{22}=\xi^2f_{11}
\eeq
where $\xi$ is the (real) ratio of the form factors $\gamma_1$ and $\gamma_2$,
whose value is obtained by
$\xi^2=|f_{12}|^2/|f_{11}|^2=\sigma_{inel}/\sigma_{el}$. All parameters are
then
fixed by the value of the total and the elastic nucleon-nucleon cross
sections, namely by
the imaginary part and by the modulus of $f_{11}$. The resulting expression of
the single
rescattering correction to the forward amplitude is then given by Eq.(25)
with $f=f_{11}$ and $\gamma=\gamma_1$.

With a larger number of rescatterings,  our model  with the choice (23)
generates  amplitudes which are different and essentially simpler
as compared to the two-state Glauber model.
This raises the  problem of unitarity in our approach.

A simple way to satisfy unitarity to all orders of rescattering in our
model,
is to ensure that unitarity is fulfilled for the individual scattering
matrices (17). Namely one has to enforce
\beq
 2{\Re}{\rm e}\Big(J(z)\Gamma(b)\Big)\geq \Big|J(z)\Gamma(b)\Big|^2
\eeq
at all values of $z$ and $b$. At first sight this condition is
not so easy to fulfill, since it involves the real part of the nucleon-nucleon
scattering amplitude on the
left-
hand side, which may have different signs at different energies.
However we can  satisfy (29) if we assume that $\tau(v)$ is
itself an analytic function in the lower half plane. Rotating the
contour in (22) to pass along the negative imaginary axis, we can rewrite
(22) as
\beq
J(z)=\theta(z)\int_0^{+\infty}
dv\tau_1(v)\left(1-\exp\left(-\frac{xmvz}{Q^2}\right)\right)
\eeq
where $\tau_1(v)=-i\tau(-iv)$. Now it is sufficient to require that
$\tau_1$ is positive to have $0\leq J(z)\leq 1$, in which case (29)
is satisfied provided the amplitude $\Gamma$ is itself unitary.

Making again the simplest choice $\tau_1(v)=\delta(v-M^2)$, we obtain
a purely real  $J(z)$
\beq
J(z)=\theta(z)
\Big[1-\exp\left(-\frac{z}{l(Q^2)}\right)\Big]
\eeq
where the formation length $l(Q^2)$ is defined by Eq. (24).
In the past the formation length was often introduced into the rescattering
picture  in a straightforward manner, essentially by changing the
function $\theta(z)$
by $\theta(z-l)$ in the rescattering matrix.
In our approach, with (29),
we also find a real damping factor in the rescattering matrix, which however
has a much softer
behavior and vanishes at high $Q^2$ only as $\sim 1/Q^2$:
\beq
J(z)\simeq \theta(z)\frac{xmz}{Q^2}\int dv v\tau_1(v)=
\theta(z)\frac{xmzM^2}{Q^2},\ \ Q^2\rightarrow\infty
\eeq
where $M^2=<v>$ is the average excitation mass squared (with $m^2$ subtracted).
Of course, (32) is true only if this average exists, that is if the integration
over $v$ in the first expression on the right-hand side converges. If
not,
 the  vanishing of  $J(z)$  at large $Q^2$ is slower.

Assuming (32) we find that the
propagation of the ejectile between any two points
in the nucleus along the $z$ axis, gives a
small factor $\sim mM^2R_A/Q^2$. Eq. (15) then tells us that the
amplitude with
$n$ rescatterings
behaves as $1/l^{2n}$, that is, as $1/Q^{4n}$. The leading
rescattering correction will come from the single rescattering term and
it is of order $1/Q^4$. It is interesting that the $n$-fold
rescattering amplitude obtained from the consecutive rescattering diagram
(Eq. (8))
generally has a slower decrease  with $Q^2$, namely it decreases
 as $1/Q^{2(n+1)}$ (with the
exception of the single rescattering term, $n=1$, when both amplitudes
coincide).
This means that due to FFT the
 total absorptive corrections to the
structure function,
generated by the direct interaction of the incoming photon with a nucleon
as a whole, is of order $1/Q^4$,
 as compared with the plane wave impulse approximation.
It is remarkable that the contribution of the rescattering,
although also vanishing at $Q^2\rightarrow\infty$,
has a relative order of $1/Q^2$, and so it is substantially larger than
the total absorptive corrections.
This follows from our expression (20) for the discontinuity. Evidently in the
limiting case $Q^2\rightarrow\infty$ only the two first terms survive,
which correspond to the cut of the nucleon propagators (which does not
correspond properly to a
rescattering, but rather to an interference term).
Using (31) we  find in the limit $Q^2\rightarrow\infty$
\[
\frac{F_{2}^{N/A,(1)}}{F_{2}^{N/A,(0)}}=
-{1}{2}(A-1)mxM^2\sigma^{tot}
\frac{M^2}{Q^2\langle 1\rangle_1}\]\beq
\langle(U(b,z_1)+U(b,z'_1)-z_1T(b,z_1)-z'_1T(b,z'_1))\rangle_1
\eeq
where
\beq
U(b,z)=\int_{z}^{+\infty}dz'z'\rho(b,z')
\eeq
and $\sigma^{tot}=2{\Im}{\rm m}f$ is the total cross-section for the
NN interaction.

\section{FSI for the deuteron target}

The deuteron structure function may be written as follows
\begin{equation}
F_{2}^d=F_{2}^{N/d,(0)}+F_{2}^{N/d,(1)}+F_{2}^{N^*/d,(1)}
\label{effe2}
\end{equation}

\noindent where $F_{2}^{N/d,(0)}$ is the expression obtained in impulse
 approximation, $F_{2}^{N/d,(1)}$ the
contribution to the deuteron structure function in presence of FSI and with a
proton in the final state, while all other contributions to the structure
function are represented by $F_{2}^{N^*/d,(1)}$. The structure function is
obtained by
working out the imaginary parts of the
the forward virtual photon -
deuteron amplitude.
For a deuteron target the amplitude without rescattering is given by:
\beq
{\cal A}^{(0)}=(1/2)i\gamma^2(Q^2)x^2m\int d^2bdz_1dz'_1\psi(b,z_1)\
\psi(b,z'_1)
\theta(z'_1-z_2)\exp(i\Delta(z'_1-z_1))
\eeq
where $\psi(b,z)$ is the deuteron wave function and, to keep into account
virtual photon finite
energy effects, $\Delta=Q^2(1-x)/(2q_zx)$.
The discontinuity of this amplitude, corresponding to the cut
active nucleon line, gives the contribution to  the inclusive
deuteron structure function generated by the production of a fast
nucleon $F_{2}^{N/d}$.
We  find from (36)
\beq
F_{2}^{N/d,(0)}(x,Q^2)=\pi x^2m\gamma^2(Q^2)
\int d^3k\phi^2(k)
\delta\bigr(k_z-Q^2(1-x)/(2q_zx)\bigl)
\eeq
 $\phi(k)$ being the deuteron  wave function in  momentum space.

The amplitude  with a single rescattering is written as
\[
{\cal A}^{(1)}=-(1/2)\gamma^2x^2m
\int dz_1dz'_1d^2b\psi(b,z_1)i\Gamma(b)\psi(b,z'_1)J(-z_1)J(z'_1)
\exp(i\Delta(z'_1-z_1))\]
where $b$ is the distance
between the proton and the neutron in transverse space. More explicitly in the
two-channel model one has:

\beq
{\cal A}^{(1)}=(1/2)\gamma^2x^2m\int d^2b i\Gamma(b)\bigl[X(b,x,Q^2)\bigr]^2
\eeq
where

\begin{eqnarray}
X(b,x,Q^2)&=&i\int dz\psi(b,z)J(-z)\exp(i\Delta z)\nonumber\\
&=&\int\frac{d^3k}{(2\pi)^{3/2}}\phi(k)e^{i{\bf k}_t{\bf
b}}\Biggl[\frac{1}{k_z-\Delta-i0}-\frac{1}{k_z-\Delta+\frac{1}{l}-i0}\Biggr]
\end{eqnarray}

The corresponding  cross-section
to produce a fast nucleon has two different contributions:
from the cut of the amplitude $\Gamma$ and from the cut of the nucleon
propagators.
The sum of the two discontinuities from the cut nucleon propagators
is
given by
\beq
{\rm Disc}_1{\cal A}^{(1)}=ix^2m\gamma^2\int
d^2bY(b,x){\Re}{\rm e}\bigl[i\Gamma(b)X(b,x,Q^2)\bigr]
\eeq
where
\beq Y(b,x)=
\int\frac{d^3k}{(2\pi)^{3/2}}\phi(k)e^{i{\bf k}_t{\bf
b}}2\pi\delta(k_z-\Delta)
\eeq
and, in the Bjorken limit, it is a real function independent of  $Q^2$.

As for the discontinuity corresponding to a cut across the
rescattering
blob $\Gamma$, since we are interested in the contribution of the scattered
nucleon to the inclusive structure function, only the
elastic part of the unitarity sum over the intermediate states has to be
retained. We obtain:
\beq
{\rm Disc}_2{\cal A}^{(1)}=i(1/2)x^2m\gamma^2\int d^2b|\Gamma(b)|^2|X(b,x,Q^2)|^2
\eeq

The  contribution to the inclusive  deuteron structure function due to the
nucleon rescattering in the final state, is given by the sum of the two
discontinuities (40) and (42) divided by $i$
\beq
F_{2}^{N/d,(1)}=-i( {\rm Disc}_1{\cal A}^{(1)}+{\rm Disc}_2{\cal A}^{(1)})
\eeq

Note that at low energy, when no elastic channels are open and
$\sigma_{tot}=\sigma_{el}$, one has
\begin{eqnarray}
F_{2}^{N/d,(1)}&=&x^2m\gamma^2\int
d^2b\Biggl\{-2{\Im}{\rm
m}\Gamma(b)+\frac{|\Gamma(b)|^2}{2}\Biggr\}\frac{\bigl[Y(b,x)\bigr]^2}{4}
\nonumber\\
&=&x^2m\gamma^2\int d^2b\Biggl\{-{\Im}{\rm m}\Gamma(b)\Biggr\}\frac{\bigl[Y(b,x)
\bigr]^2}{4}=2
{\Im}{\rm m}{\cal A}^{(1)}
\end{eqnarray}

The only contributions to the imaginary part of the forward amplitude is given,
in this case,
by the the two discontinuities (40) and (42) where only the elastic
intermediate state is present.
At higher energies the inelastic
channels become more and more important. The effect is to add further
contributions to the imaginary part of the forward amplitude. As it may be seen
by looking at the behavior of $X$, Eq.(38), as a function of the formation
length
$l$, the additional
contributions give a small correction at low $Q^2$ (small $l$) while they tend
to cancel completely the elastic contribution at large $Q^2$ (large $l$).

To study quantitatively the behavior of the ratio which characterizes
the strength
of the FSI, viz.
\beq
R_N(Q^2)= 1 +
\left(\frac{F_{2}^{N/d,(1)}(x,Q^2)}
{F_{2}^{N/d,(0)}(x,Q^2)}\right)_{x=1}
\eeq
one has to specify the value of the mass parameter $M^2$.
Its meaning is that of the squared  average excitation mass
of the ejectile: $M^2=(m^*)^2-m^2$, where $m^*$ is the
average mass of the ejectile. Previous calculations, based on
the coupled-channel
Glauber formalism have shown that $m^*$ lies between  the lowest N$^*$
resonance mass 1.44 GeV and the average continuum mass 2.4 GeV, thus  a reasonable
value could be 1.8 GeV [10,14,16].
Predictions for $R_N(Q^2)$
for these values of $m^*$ are shown in figure 2 at $x=1.$
For the deuteron wave function we have  used the parameterization of [18],
corresponding to a realistic nucleon-nucleon interaction, and for
the nucleon-nucleon amplitude we have used both the experimental data
[19] and the results of the partial wave analysis[20]. As a comparison we also
show the pure Glauber predictions, which
correspond to a very large value of $m^*$. The results in the pure Glauber case
are in agreement with those obtained in Ref.[21], where both the
Reid Soft Core and the Bonn deuteron wave functions have been used and
interference effects between deuteron $S$ and $D$ waves have
been taken into account explicitly.
In all cases the FSI are found to be small,
as to be expected due to the large deuteron size. As for the
behavior with $Q^2$, the values of both the threshold and the rate at which
$R(Q^2)$ goes to one depend on the value chosen for $m^*$, the threshold
growing and the rate diminishing with $m^*$. For the
value $m^*$=1.8 GeV and at $x=1$, the FSI changes in a sizable way when
$Q^2$ is of the order of 10 (GeV/c)$^2$, which agrees
with the conclusions inferred from the conventional approach to color transparency.

The FFT approach allows one also to calculate the total structure function
at $x\sim 1$, provided that
the initial interaction involves a proton as a whole.
By looking at the imaginary part of the rescattering amplitude one can in
fact work out the FSI for the total
structure which vanishes faster, like $1/Q^4$, as compared to $R_N(Q^2)$.
The behavior is illustrated in figure 3
where we show
\beq
R^{tot}(Q^2)=1 + \left(\frac{F_{2}^{N/d,(1)}(x,Q^2)+F_{2}^{N^*/d,(1)}(x,Q^2)}
{F_{2}^{N/d,(0)}(x,Q^2)}\right)_{x=1}
\eeq
and where $F_{2}^{N/d,(1)}+F_{2}^{N^*/d,(1)}$ is evaluated by taking twice the
imaginary part of (38). Looking at the
continuous curve, corresponding to an excitation mass $m^*=1.8$ GeV,
one observes that the threshold at which the FSI
starts to vanish is practically the same as for the proton production, while
the effect of FSI is sizably smaller in this case.

\section{Conclusions}
We have studied the FFT effects by introducing the dependence on the
virtualities into the elementary amplitudes.
Two options have been considered for generalizing the standard
on-shell Glauber picture to take into account the virtuality of the ejected
nucleon: the Feynman diagram and $S$-matrix factorization approaches.
The latter choice seems to be more convincing, since it preserves
both the overall unitarity and the AKG cutting rules. The single
rescattering term is however the same in both approaches,
so that its calculation seems to be reliable. Moreover the
single rescattering term can  be understood also in terms of the conventional
multichannel picture of the FSI, showing in this way that the
present approach is essentially equivalent to the conventional one
 at the single rescattering level.
In addition to a better
understanding, which is gained when a given mechanism of interaction can be
described from different perspectives, an advantage of the actual
approach lies in its far simpler
implementation, as compared with the standard multichannel description of the
FSI.

Numerical estimates are made for the deuteron target, where all FSI are
described
by the single rescattering term. Our result is that for the
q.e. $(e,e')$ reaction
the FFT effects become clearly visible at rather high values of $Q^2$, namely
for $Q^2\simeq10 (GeV/c)^2$, in accordance with the conclusions drawn within the
CT approach.

\vskip.15in
{\bf Acknowledgments}
\vskip.15in
This work was partially supported by the Ministero dell'Universit\`a e della
Ricerca Scientifica (MURST) through the funds COFIN99.
M.A. Braun is deeply thankful to the Universities of Perugia and Trieste and
to  INFN, Sezioni di Perugia and Trieste  for their hospitality and financial support.
The authors gratefully acknowledge discussions with L. A. Kaptari.

\vskip.15in
{\bf Appendix. The rescattering amplitude of Fig. 1}
\vskip.15in
One standardly starts  by the integrations over the zero components of the
momenta.
Since the poles coming from the  propagators of the active nucleon all lie in the
upper half-plane, one can integrate over $k_{i0}$ or $k'_{i0}$,
$i=2,...A$, just taking the residue at the pole of the corresponding propagator
$P(k_i)$ or $P(k'_i)$.
The two propagators of the active nucleon in the initial and final state
together with the factors $i\Phi(k_i)i\Phi(k'_i)$ then combine
into a product of two nuclear wave functions
\[(2m(2\pi)^3)^{A-1}\phi(k_i)\phi(k'_i)\]

Passing to the coordinate space wave functions $\psi(r_i)$ one then
integrates over the transverse momenta.  It is quite
trivial, since all interactions, as well as the left propagators of the active
nucleon do not depend on the small transverse momenta. So all the
dependence is in the exponentials.
One chooses one of
the spectator momenta
(say the $A$'s) and $k'_1$ as
dependent variables. Integration will then go over $A-1$
momenta $k_1, k_2,...k_{A-1}$ and $n$ momenta $k'_2,k'_3,...k'_{n+1}$.
The latter $n$ integrations can be substituted by $n$ integrations over the
transferred momenta $q_2,q_3,... q_{n+1}$.
One obtains
\[
i{\cal A}=\int\prod_{j=1}^{A-1}\frac{dk_{zj}}{2m(2\pi)}
\prod_{j=2}^{n+1}\frac{dq_{zj}}{2m(2\pi)}
P(k_1^{(1)})\prod_{j=2}^{n+1}P(k^{(j)}_1)if_j\]\[
i\gamma(k_1,q) i\gamma(k'_1,q)(2m)^{A-1}
\prod dz_i\exp(-i\sum k_{zj}z_j)
\prod dz'_i\exp(i\sum k'_{zj}z'_j)\]\beq
d^2b_1\prod_{j=n+2}^{A-1}d^2b_j
\psi(b_1=b_2=b_3+...b_{n+1},b_j; z_i)
\psi(b_1=b_2=b_3+...b_{n+1},b_j; z'_i)
\eeq

Now one integrates over the longitudinal momenta.
Evidently the
integrand does not depend on the $k_z$ of the spectators. So these
interactions are done trivially and
 convert the double integration over $z$  into a single one
for the spectators.
Together with the integration over their transverse coordinates this
turns the product of the wave functions into the nuclear $\rho$-matrix
for $n+1$ nucleons taking part in the interaction.
One is left with the $2n+1$ integrations over the  $z$ components of $k_i$,
$i=1,2,...n+1$ and $q_i$ $i=2,3,..n+1$.
All propagators of the active nucleon and also
the amplitudes $ia_j$ depend only on $v_i$, $i=1,...n+1$.
So it is convenient to pass from the $n+1$ variables
$k_{1z},q_{2z},...q_{n+1,z}$ to  variables $v_i$, $i=1,...n+1$.
The rest integration variables are $k_{2z},....k_{n+1,z}$.
The dependence on them is concentrated in the exponentials so that
 integration over them turns the double integrations over $z$
into a single one for $z_j$ with $j=2,3,...n+1$.
The left integrations are over $v_{j}$ which are done as explained
in Section 2.

The standard Glauber model corresponds to an
approximation
 in which  the dependence of the amplitudes
$f$ on their virtualities is neglected and thus the form-factor $F(v)=1$
and
\beq
J(z)=\theta(z)
\eeq
The integrations in (8) become
\[
\int_{z_1}^{z_{n+2}}dz_{n+1}\int_{z_1}^{z_{n+1}}dz_n....
\int_{z_1}^{z_3}dz_2\rightarrow
(1/n!)\int_{-\infty}^{+\infty}dz_{n+2}\int_{-\infty}^{z_{n+2}}
\prod_{j=2}^{n+1}\int_{z_1}^{z_{n+2}}dz_j
\]
where one uses the symmetry of the integrand in the variables
$z_2,z_3,...z_{n+1}$.
Each integration converts $\rho(b,z)$
 it into the
profile function between the two longitudinal points (12) and we get (11).

Various discontinuities of the amplitude (8) correspond graphically
to various cuts of the diagram shown in Fig. 1. Evidently there are
two possibilities and correspondingly two possible types of
discontinuities. The cut may pass through a fast nucleon line.
The relevant discontinuity then corresponds to an intermediate state
of a fast nucleon and the nucleus debris. Alternatively the cut may pass
through an amplitude $f$. Then the intermediate state
consists of an arbitrary ejectile state plus the nucleus debris .
 If we are interested only in the states
with one fast nucleon plus the nucleus debris, we have to select
only elastic intermediate states in the cut amplitude $f$.

Technically the discontinuity is obtained by making  substitutions
\beq \frac{1}{-v-i0}\rightarrow 2\pi i\delta(v)
\eeq
for the cut fast nucleon propagator or
\beq
f\rightarrow 2i{\Im}{\rm m}f=i\sigma
\eeq
for the cut amplitude. In the latter case only the elastic part of the
contribution to the cross-section should be taken if one is only
interested in the states with one fast nucleon. Also all the parts of
the amplitude to the right of the cut should be taken complex conjugate.

When the discontinuity passes through the fast nucleon line connecting the
points $z_k$ and $z_{k+1}$ we should make the substitution
\[J(z_{k+1}-z_k)\rightarrow 1\]
In all $J$'s to the right of the cut we have to change the sign of the $i0$
in the denominator, so that for $j>k$
\beq J(z_{j+1}-z_j)\rightarrow -J^*(z_j-z_{j+1})
\eeq
Finally of $n$ amplitudes $n-k+1$ have to be taken complex conjugate.
As a result, the total discontinuity corresponding to cut fast
nucleon lines is given by the integral (8) in which
\[I(z_1,....z_{n+2})\equiv f^n\prod_{j=1}^{n+1}iJ(z_{j+1}-z_j)\]
is substituted by
\beq
I_1(z_1,...z_{n+2})=
\sum_{k=1}^{n+1}
\prod_{j=1}^{k-1}ifJ(z_{j+1}-z_j)\prod_{k+1}^{n+1}
(if)^*J^*(z_j-z_{j+1})
\eeq

Now  let
the cut pass through the
$k$th rescattering amplitude. Then the latter should be changed
according to (49).
For $j>k$ again one should make the substitution (50) and take all the
amplitudes conjugate. Thus the second type of discontinuities will be
given by the integral (8) with $I(z_1,...z_{n+2})$ substituted by
\beq
I_2(z_1,...z_{n+2})=i\frac{\sigma^{el}}{ff^*}
\sum_{k=2}^{n+1}
\prod_{j=1}^{k-1}ifJ(z_{j+1}-z_j)\prod_{k}^{n+1}
(if)^*J^*(z_j-z_{j+1})
\eeq

It is instructive to see how the found discontinuities transform in
the case when the amplitudes do not depend on their virtualities
and $F(v)=1$.
For the discontinuity (51) we then get the $z$
integrations
\[\int_{z_1}^{\infty}dz_k\int_{z_1}^{z_k}dz_{k-1}...\int_{z_1}^{z_3}dz_2
\int_{z_{n+2}}^{\infty}dz_{k+1}\int_{z_{n+2}}^{z_{k+1}}dz_{k+2}...
\int_{z_{n+2}}^{z_n}dz_{n+1},\]
which using the symmetry of the integrand can be transformed into
\beq
\frac{1}{(k-1)!(n-k-1)!}\int_{z_1}^{\infty}\prod_{j=2}^{k}dz_j
\int_{z_{n+2}}^{\infty}\prod_{j=k+1}^{n+1}dz_j\eeq
Doing the integrations we obtain a product
\[\frac{1}{(k-1)!(n-k+1)!}[ifT(b,z_1)]^{k-1}[(ifT(b,z_{n+2})^*]^{n-k+1}\]
Summation over $k$ and $n$ gives
\beq
{\rm Disc}_1{\cal A}(b,z,z')=iA\Big(1+ifT(b,z)-if^*T(b,z')\Big)^{A-1}
\eeq
where $T(b,z)=T(b,\infty,z)$.

For the discontinuity (52) the $z$ integrations
are
\[\int_{z_>}^{\infty}dz_k\int_{z_1}^{z_k}dz_{k-1}...\int_{z_1}^{z_3}dz_2
\int_{z_{n+2}}^{z_k}dz_{k+1}\int_{z_{n+2}}^{z_{k+1}}dz_{k+2}...
\int_{z_{n+2}}^{z_n}dz_{n+1}\]
where $z_>=\max (z_1, z_{n+2})$
Fixing $z_k\equiv\zeta$ and again using the symmetry of the integrand in the
two groups of the left variables we arrange  the integrations as in (53)
and doing them get
\[ \frac{1}{(k-2)!(n-k+1)!}\int_{z_>}^{\infty}d\zeta\rho(b,\zeta)
[ifT(b,\zeta,z_1)]^{k-2}[(ifT(b,\zeta,z_{n+2})^*]^{n-k+1}\]
After summation over $k$ and $n$
the part due to cut amplitudes is found to be
\beq
{\rm Disc}_2{\cal A}(b,z,z')=iA(A-1)\sigma^{el}\int_{z_>}^{\infty}
d\zeta\rho(b,\zeta)\Big(1+ifT(b,\zeta,z)-if^*T(b,\zeta,z')\Big)^{A-2}
\eeq
with $z_>=\max (z,z')$.
Using
\[
\frac{\partial}{\partial\zeta}
\Big(1+ifT(b,\zeta,z)-if^*T(b,\zeta,z')\Big)^{A-1}=\]\beq
-(A-1)\sigma^{tot}\rho(b,\zeta)
\Big(1+ifT(b,\zeta,z)-if^*T(b,\zeta,z')\Big)^{A-2}
\eeq
one can do the integration over $\zeta$ in (55):
\[
{\rm Disc}_2{\cal A}(b,z,z')=-iA\frac{\sigma^{el}}{\sigma^{tot}}
\Big[\Big(1+ifT(b,z)-if^*T(b,z')\Big)^{A-1}-\]\beq
\Big(1+ifT(b,z_>,z)-if^*T(b,z_>,z')\Big)^{A-1}\Big]
\eeq
The sum of (54) and (57) gives the total discontinuity
\[
{\rm Disc}{\cal A}(b,z,z')=iA\left(1-\frac{\sigma^{el}}{\sigma^{tot}}\right)
\Big(1+ifT(b,z)-if^*T(b,z')\Big)^{A-1}+\]\beq
A\frac{\sigma^{el}}{\sigma^{tot}}
\Big(1+ifT(b,z_>,z)-if^*T(b,z_>,z')\Big)^{A-1}
\eeq

One immediately notes that this discontinuity (and the corresponding
proton production probability) is different from the one obtained by
squaring the non-forward production amplitudes for the process
$(e,e')$ calculated in the Glauber approach and summing over all
final states. The latter is easily obtained as [16]
\beq
{\rm Disc}{\cal A}(b,z,z')=iA\Big(1+ifT(b,z)-if^*T(b,z')+
\sigma^{el}T(b,z_>)\Big)^{A-1}
\eeq
which, for a source not extended in $z$ so that $z=z'$ transforms into
\beq
{\rm Disc}{\cal A}(b,z,z)=iA\Big(1
-\sigma^{in}T(b,z)\Big)^{A-1}
\eeq
with a clear probabilistic interpretation.
The reason for this difference
has long been known: the Glauber -Gribov picture of consecutive
rescatterings corresponding to Fig. 1 is effectively valid for the
amplitude itself but not for its discontinuities, due to the wrong
space-time picture inherent in it.

If one is interested in the distribution of the produced fast nucleons
in the
momentum space then the discontinuities taken in Section 2 have to be further
specified.
It is quite simple to do it if the cut
passes through one of the rescattering amplitudes. Then the inclusive
cross-section of interest
 is obtained by substituting the cut amplitude by the relevant
inclusive cross-section for the collision of the active nucleon
(momentum $k_1+q$) with a nucleon at rest.

The contribution to the inclusive cross-section
coming from a cut ($i$th) propagator is a bit more complicated.
Now one has to substitute the propagator by
\[(2\pi)^4\delta(v_i)\delta^3(k^{(i)}-l)=2\pi\delta(v_i)
\int d^3R\exp i{\bf (k^{(i)}-l)R}\]
The additional exponential function will somewhat change our
derivation.

In the transverse part of the exponent
 apart from  $-il_TR_T$, we have  additional terms
$ik_{1T}R_{T}$, which will shift the argument of the corresponding $\delta$
 function
by $-R_T$, and a term
$-iq_{jT}R_T,\ \ j\leq i$
which will also shift the arguments in the corresponding $\delta$ functions
 by $R_T$.
As a result   the transverse coordinates
in the nuclear wave functions in  Eq. (44) become
$b_j=b'_j=b_1$ for $j=2,...i$; $b_j=b'_j=b'_1=b_1-R_T$ for $j>i$.
Shifting the $b_{1}$ integration by $R_T$ we make them
$b_j=b'_j=b_1+R_T$ for $j=2,...i$; $b_j=b'_j=b'_1=b_1$ for $j>i$.

As to the longitudinal part, the additional exponent in terms of $v_i$
has the form
\[-iZ((xm/Q^2)v_i+\Delta+q_z-l_z)\]
However $v_i=0$ so that after integration over $Z$ we obtain a factor
$
2\pi\delta(l_z-q_z-\Delta)
$
Thus the observed fast nucleon carries the longitudinal momentum of the
initial photon shifted by $\Delta$.
No other effect is introduced by the longitudinal exponent.

So in the end the inclusive cross-section to produce a fast nucleon
with the momentum $\bf l$, corresponding to the cut $k$-th line will be given
by the expression
\[
I_1^{(i)}({\bf l})=i\gamma^2(Q^2)
(2\pi)\delta(l_z-q_z-\Delta)
\frac{x^2m}{2}\frac{A!}{(A-n-1)!}\]\[
\int d^2bd^2b'\exp (i l_{\bot}(b-b'))\prod_{j=1}^{n+2}dz_j
\exp (i\Delta(z_{n+2}-z_1))\rho(bz_1|b'z_{n+2})\]\[
\]\beq
\prod_{j=2}^{k}if\rho(bz_j)J(z_j-z_{j-1})\prod_{k+1}^{n+1}
(if)^*\rho(b'z_j)J^*(z_j-z_{j+1})
\eeq

\vskip.15in
{\bf References}
\vskip.15in

\noi [1] A.H.Mueller, in {\it Proc. 17th Recontre de Moriond}, ed. J.Trahn Thanh Van,
Frontieres, Gif-sur-Ivette, 13 (1982);

\noi [2] S.J.Brodsky, in {\it Proc. 13th Int. Symp. on Multiparticle Dynamics},
ed W.Kittel, W.Metzger and A.Stergiou, World Sci., Singapore, 963 (1982);

\noi [3] L.Frankfurt, G.A.Miller and M.Strikman, {\it Nucl. Phys.}
{\bf A555}, 752 (1993);

\noi [4] N.N.Nikolaev, {\it Pis'ma JETP} {\bf 57}, 82 (1993);

\noi [5] O. Benhar, S. Fantoni, N.N. Nikolaev, J.Speth, A.A. Usmani and B.G.
Zakharov, {\it J. Exp. Theor. Phys.} {\bf 83}, 1063 (1996)
[{\it Zh. Eksp. Teor. Fiz.} {\bf 110}, 1933 (1996)];

\noi [6] T. Iwama, A. Kohama and K. Yazaki, {\it Nucl. Phys.} {\bf A627}, 620
(1997);

\noi [7] V.N.Gribov, {\it Sov. Phys. JETP} {\bf 29}, 483 (1969);
{\bf 30} (1970) 709; L. Bertocchi, {\it Nuovo Cimento} {\bf 11A}, 45 (1972);

\noi [8] S.J.Brodsky and A.H.Mueller, {\it Phys. Lett.} {\bf B206}, 285 (1988);

\noi [9] B.Z.Kopeliovich and B.G.Zakharov, {\it Phys. Rev.} {\bf D44}, 3466
(1991);

\noi [10] A.Bianconi, S.Boffi and D.E.Kharzeev, {\it Nucl. Phys.} {\bf A565}, 767
(1993);

\noi [11] L.V. Gribov, E.M. Levin and M.G. Ryskin, {\it Phys. Rep.} {\bf 100}, 1
(1983);

\noi [12] V.V. Anisovich, L.G. Dakhno and M.M. Giannini, {\it Phys. Rev.} {\bf
C49}, 3275 (1994);

\noi [13] J.H. Weis, {\it Acta Physica Polonica} {\bf B7}, 851 (1976);

\noi [14] V. Abramovskii, V.N. Gribov and O.V. Kancheli, {\it Yad. Fiz.}
{\bf 18}, 595 (1973) [{\it Sov. J. Nucl. Phys.} {\bf 18}, 308 (1974) ];

\noi [15] N.N. Nikolaev, J. Speth and B.G. Zakharov, {\it J. Exp. Theor. Phys.}
{\bf 82}, 1046 (1996) [{\it Zh. Eksp. Teor. Fiz.} {\bf 109}, 1948 (1996)];

\noi [16] L. Bertocchi and D. Treleani, {\it Nuovo Cimento} {\bf A34}, 193 (1976);

\noi [17] L. Frankfurt, W.R. Greenberg, G.A. Miller and M.
Strikman, {\it Phys. Rev.} {\bf C46}, 2547 (1992);

\noi [18] C. Ciofi degli Atti and S. Simula, {\it Phys. Rev.} {\bf C53}, 1689
(1996);

\noi [19] A. Baldini {\it et al} in 'Total cross Section for Reactions of High
Energy particles", Ed. H. Schopper, Springer Verlag, Berlin, 1987;

\noi [20] R.A. Arndt {\it et al}, "(SAID) Partial-Wave Analysis Facility",
http://said.phys.vt.edu/;

\noi [21] C. Ciofi degli Atti, L.P. Kaptari and D. Treleani, in preparation.

\begin{figure}
\centerline{
\epsfysize=23cm \epsfbox{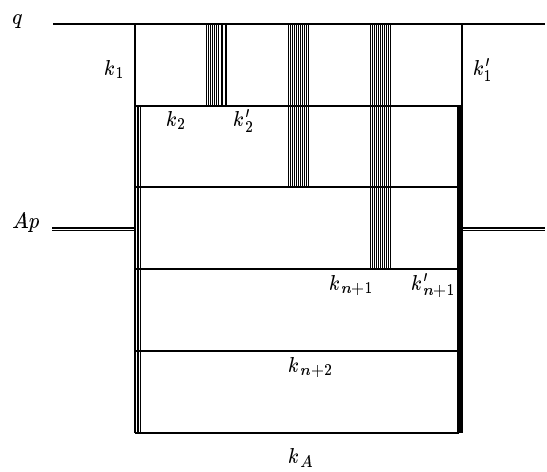}
}
\vspace{-4 cm}
\caption[ ]{The forward scattering amplitude.}
\label{ampl}
\end{figure}

\begin{figure}
\vspace{2 cm}
\centerline{
\epsfysize=15cm \epsfbox{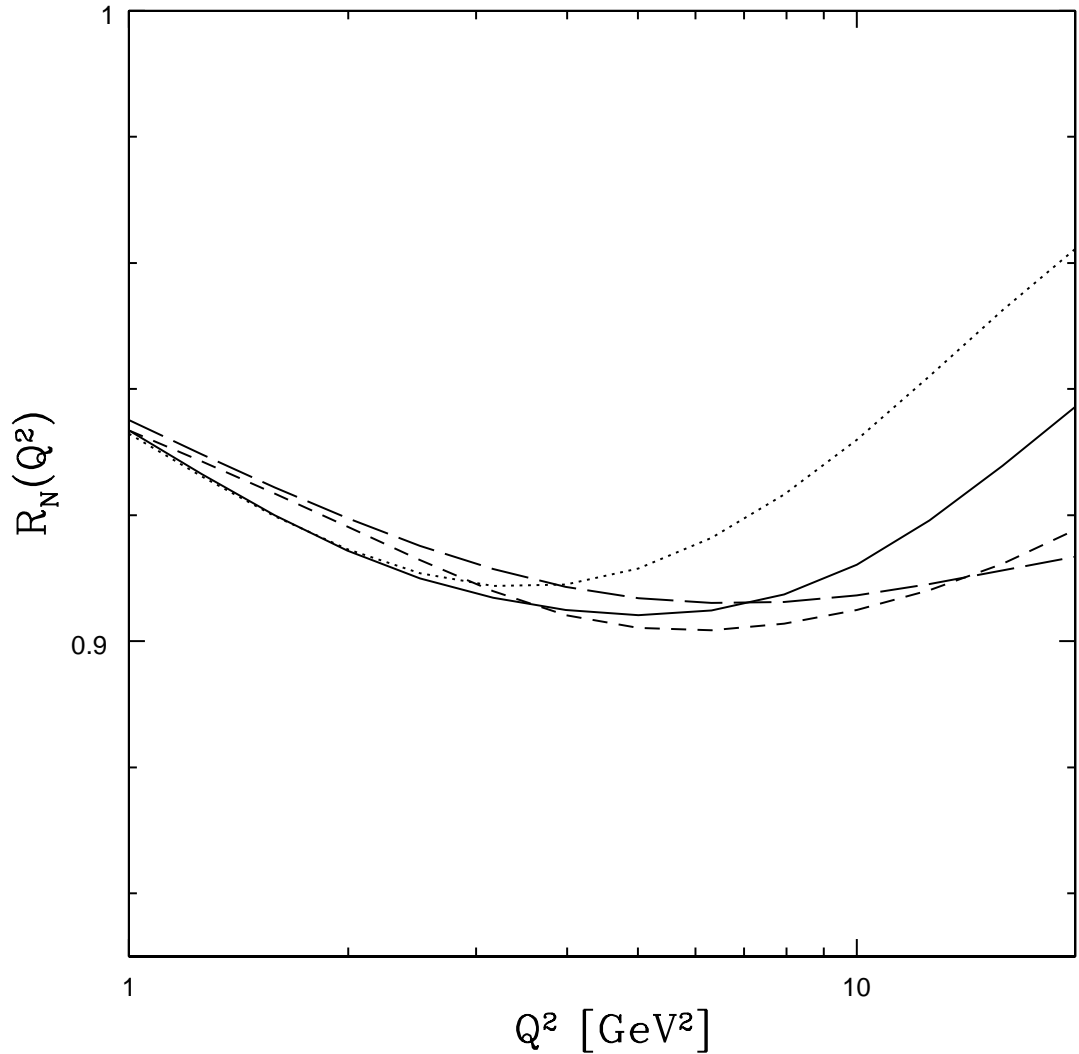}
}
\vspace{1 cm}
\caption[ ]{Values of $R_N(Q^2)$ (eq. (45)) at $x=1$ for the deuteron target
with different choices of the
excited nucleon mass: $m^*=1.44$ (GeV) dotted line, $m^*=1.8$ (GeV)
continuous line, $m^*=2.4$ (GeV) short-dashed line. The long-dashed line
corresponds to the standard Glauber result, where no dependence of the amplitude on the
virtuality of the external lines is taken into account.}
\label{fft1}
\end{figure}

\begin{figure}
\vspace{2 cm}
\centerline{
\epsfysize=15cm \epsfbox{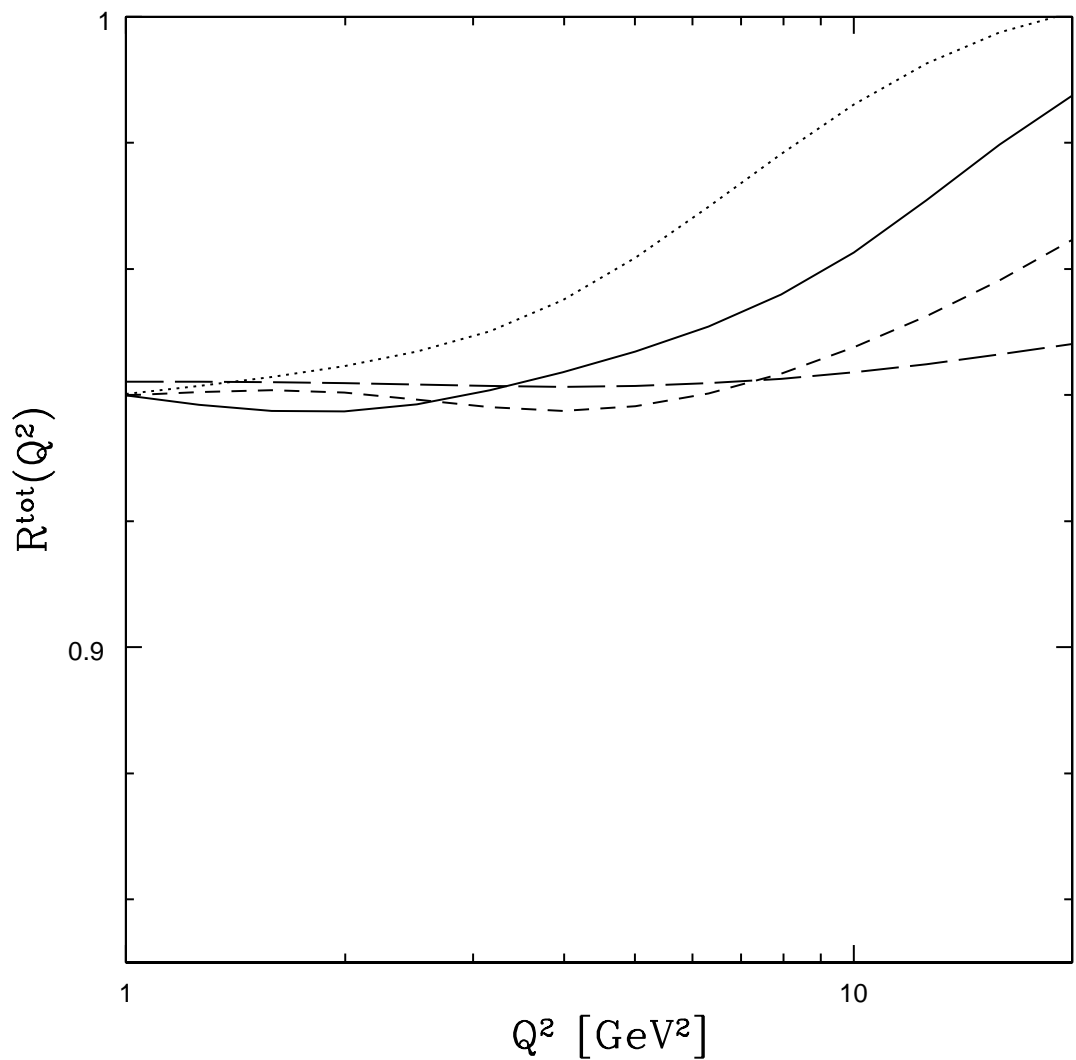}
}
\vspace{1 cm}
\caption[ ]{Values of $R^{tot}(Q^2)$ (eq. (46)) at $x=1$ for the deuteron target. The
different lines refer to the different cases described in the previous figure.}
\label{fft1}
\end{figure}

\end{document}